\documentclass[11pt]{article}
\usepackage{amsmath,amssymb,color,graphics,epsfig}

\usepackage{amsfonts}

\newcommand{\hoch}[1]{$\, ^{#1}$}

\newcommand{\be}{\begin{equation}}
\newcommand{\ee}{\end{equation}}
\newcommand{\bea}{\setlength\arraycolsep{2pt} \begin{eqnarray}}
\newcommand{\eea}{\end{eqnarray}}

\def\ft#1#2{{\textstyle{\frac{\scriptstyle #1}{\scriptstyle #2} } }}
\def\fft#1#2{{\frac{#1}{#2}}}

\def\0{{\sst{(0)}}}
\def\1{{\sst{(1)}}}
\def\2{{\sst{(2)}}}
\def\3{{\sst{(3)}}}
\def\4{{\sst{(4)}}}
\def\5{{\sst{(5)}}}
\def\6{{\sst{(6)}}}
\def\7{{\sst{(7)}}}
\def\8{{\sst{(8)}}}
\def\sst#1{{\scriptscriptstyle #1}}

\begin{document}

\begin{flushright}
\end{flushright}

\vspace{25pt}
\begin{center}
{\large {\bf Exact Black Hole Formation in Asymptotically (A)dS and Flat Spacetimes}}

\vspace{10pt}
Xuefeng Zhang\hoch{1,2} and H. L\"u\hoch{1}

\vspace{10pt}

{\it \hoch{1}Department of Physics, \hoch{2}Department of Astronomy,\\
Beijing Normal University, Beijing 100875, China}

\vspace{40pt}

\underline{ABSTRACT}
\end{center}

We consider four-dimensional Einstein gravity minimally coupled to a dilaton scalar field with a supergravity-inspired scalar potential. We obtain an exact time-dependent spherically symmetric solution describing gravitational collapse to a static scalar-hairy black hole. The solution can be asymptotically AdS, flat or dS depending on the value of the cosmological constant parameter $\Lambda$ in the potential.  As the advanced time $u$ increases, the spacetime reaches equilibrium in an exponential fashion, i.e., $e^{-u/u_0}$ with $u_0\sim1/(\alpha^4 M_0)^{1/3}$, where $M_0$ is the mass of the final black hole and $\alpha$ is the second parameter in the potential. Similar to the Vaidya solution, at $u=0$, the spacetime can be matched to an (A)dS or flat vacuum except that at the origin a naked singularity may occur. Moreover, a limiting case of our solution with $\alpha=0$ gives rise to an (A)dS generalization of the Roberts solution, thereby making it relevant to critical phenomena. Our results provide a new model for investigating formation of real life black holes with $\Lambda \geq 0$. For $\Lambda<0$, it can be instead used to study non-equilibrium thermalization of certain strongly-coupled field theory.

\vfill {\footnotesize Emails: zhxf@bnu.edu.cn\ \ \ mrhonglu@gmail.com}

\thispagestyle{empty}

\pagebreak



\newpage

{\bf Introduction}: Many subtle properties and mysteries of black holes have been revealed, largely owing to the discovery of exact solutions such as the Schwarzschild and Kerr metrics. Formation of black holes due to gravitational collapse is of great interest \cite{Joshi12,Gundlach07,Fryer11} for addressing important issues such as spacetime singularities, cosmic censorship and gravitational waves. Nevertheless, exact solutions that describe collapsing to static black holes with specific classes of matter energy-momentum tensors are quite rare \cite{Vaidya51,Husain95,Wang99}.  A most prominent one is perhaps the Vaidya metric \cite{Vaidya51}, which captures some essential features of time evolution. However, generalizing this line of work to a Lagrangian of fundamental fields seems to be quite a formidable task, and no such solutions have been previously known, to our best knowledge, in literature.

Motivated by the AdS/CFT correspondence \cite{Mald98}, there has been a resurgence of interest lately in constructing black holes in asymptotically anti-de Sitter (AdS) spacetimes. In light of this new trend, a time-dependent black hole formation can provide a useful background for certain non-equilibrium thermal systems of strongly-coupled dual gauge theories. Although analytic results can be obtained in some limiting parameter regions \cite{Bhattacharyya09}, the study of more general properties has to be conducted with numerical approach so far. (See e.g. \cite{Chesler13,Craps13}.)

In this paper, we report an exact solution describing the formation of a scalar-hairy black hole in asymptotically (A)dS or flat (Minkowski) spacetimes in four-dimensional Einstein gravity minimally coupled to a dilatonic scalar with a specific scalar potential.  We first present the local solution and show that it eventually settles down to a static black hole.  Then we study the global property for general time, and establish the initial energy-momentum tensor that kick-starts the evolution. Furthermore, we derive a limiting solution of the theory which contains the Robert solution as a special case. The paper is concluded with further discussions.

{\bf The theory:} The Lagrangian is given by
\bea
{\cal L} &=& \sqrt{-g} \left(R - \ft12(\partial\phi)^2 - V(\phi)\right),\cr
V&=& -2g^2 (\cosh\phi + 2) - 2\alpha^2 (2\phi + \phi \cosh\phi - 3 \sinh\phi)\,,
\label{lag1}
\eea
with two parameters $g^2$ and $\alpha^2$ (negative values allowed).
The scalar potential $V$ has a stationary point at $\phi=0$ with Taylor expansion
$ V=-6g^2 - g^2 \phi^2 - \ft1{12} g^2 \phi^4 -\ft{1}{30}\alpha^2 \phi^5+ \cdots$.
This potential was first introduced by Zloshchastiev \cite{Zlosh05}.  In fact, when $\alpha^2=0$, the potential with $g^2>0$ arises from ${\cal N}=4$, $D=4$ gauged supergravity \cite{Gates83}.  In this paper, we allow the parameter $g^2$, hence the cosmological constant $\Lambda=-3g^2$ as well, to be negative, zero or positive as needed. Note that solutions to (\ref{lag1}) come in pairs through the inversion $\phi\rightarrow -\phi$, $\alpha^2\rightarrow -\alpha^2$ which leaves the potential $V$ invariant.

{\bf The local solution}: Using Eddington-Finekelstein-like coordinates, our new solution reads
\bea
ds^2 &=& 2 du dr - H(r,u) du^2 + r\Big(r + q\tanh(\ft12\alpha^2 q\, u)\Big) d\Omega_{2,k}^2\,,\quad e^{\phi} = 1 + \fft{q}{r} \tanh(\ft12\alpha^2 q\, u)
\,,\cr
H &=& g^2r^2 + k -\ft12\alpha^2 q^2 + (g^2-\alpha^2)q r \tanh(\ft12 \alpha^2 q\,u)\cr
&& + \alpha^2 r^2 \Big(1 + \fft{q}{r}\tanh(\ft12\alpha^2 q\,u)\Big)
\log\! \Big(1 + \fft{q}{r}\tanh(\ft12\alpha^2 q\,u)\Big)\,.
\eea
where $d\Omega_{2,k}^2=dx^2/(1-kx^2) + (1-k x^2) dy^2$ is the metric on a 2-space of constant Gaussian curvature normalized to $k=0,\pm1$. It also contains one free parameter (a constant of integration), namely the ``scalar charge'' $q$. In addition, the solution belongs to a generalized class of Robinson-Trautman solutions \cite{Guven96}, and it is of Petrov type D with $\Psi_2$ being the only non-vanishing Weyl scalar.

{\bf Static hairy black holes}:  Assuming $\alpha^2>0$ and $q>0$, then in the static limit $u\rightarrow \infty$, we have $\tanh(\ft12\alpha^2 q\,u)\rightarrow 1$ and the solution becomes
\bea
ds^2 &=& -f dt^2 + \fft{dr^2}{f} + r(r+q) d\Omega_{2,k}^2\,,\qquad
e^{\phi}=1 + \fft{q}{r}\,,\cr
f(r)&=&g^2r^2 + k -\ft12\alpha^2 q^2 + (g^2-\alpha^2)q r+ \alpha^2 r^2 \Big(1 + \fft{q}{r}\Big)\log\! \Big(1 + \fft{q}{r}\Big)\,.\label{staticbh}
\eea
Here we have made the coordinate transformation $dt=du + dr/f$ to put the metric in Schwarzschild-like coordinates. The exact form of (\ref{staticbh}) was first obtained in \cite{Zlosh05} and proposed as a cosmological model. Discussion on the AdS case can be found in \cite{Gonzalez13}. The asymptotic behavior indicates that the mass is given by $M_0=\alpha^2 q^3/12$. The horizon is located at the largest root $r=r_0>0$ of the function $f$. Using the standard technique for computing the temperature $T$ and the entropy $S$, one can verify that $dM=TdS$ \cite{Gonzalez13}. For $g^2 \geq 0$, the function $f$ is strictly increasing to infinity or $k$.  An event horizon therefore uniquely exists iff $f(0)<0$ and $f(\infty)>0$, or more specifically, $k - \ft12 \alpha^2 q^2 <0$ with $k$=1 if $g^2=0$. For $g^2<0$, extremal limit may occur when the cosmological and black hole horizons coincide, which imposes certain bounds on physical parameters in cosmological spacetimes \cite{Zlosh05}.  It is worth remarking that for a given mass, the theory (\ref{lag1}) also admits a Schwarzschild black hole satisfying the same first law of thermodynamics but with different temperature and entropy, therefore making black holes hairy.

{\bf Time evolution and singularity:}  To study properties of the time evolution, it is instructive to use the luminosity coordinate $R$ which is the radius of $d\Omega_{2,k}^2$.  Picking the positive root for $r$, i.e.
\be
r=\ft12 \left(\sqrt{4R^2 + q^2 \tanh^2(\ft12\alpha^2 q\,u)} - q\tanh(\ft12\alpha^2 q\, u)\right),
\ee
we can rewrite the solution as
\bea
ds^2 &=& 2h\,du\, dR - \tilde H du^2 + R^2 d\Omega_{2,k}^2\,,\qquad
h= \fft{2R}{\sqrt{4R^2 + q^2 \tanh^2(\ft12\alpha^2 q\,u)}}\,,\cr
\tilde H &=& g^2 R^2 + k -
\fft{\alpha^2 q (4R^2 + q^2) \tanh(\ft12\alpha^2 q\,u)}{2 \sqrt{4R^2 + q^2 \tanh^2(\ft12\alpha^2 q\,u)}}\cr
&&+\alpha^2 R^2 \log\!\left(\fft{\sqrt{4R^2 + q^2 \tanh^2(\ft12\alpha^2 q\,u)} + q\tanh(\ft12\alpha^2 q\, u)}{\sqrt{4R^2 + q^2 \tanh^2(\ft12\alpha^2 q\,u)} - q\tanh(\ft12\alpha^2 q\, u)}
\right)\,,\cr
e^{\phi} &=& \fft{\sqrt{4R^2 + q^2 \tanh^2(\ft12\alpha^2 q\,u)} + q\tanh(\ft12\alpha^2 q\, u)}{\sqrt{4R^2 + q^2 \tanh^2(\ft12\alpha^2 q\,u)} - q\tanh(\ft12\alpha^2 q\, u)}
\,.
\eea
For large $R$, the metric functions behave as
\bea
h &=& 1 - \fft{q^2\tanh^2(\ft12\alpha^2 q\,u)}{8R^2} + {\cal O}(R^{-4})\,,\cr
\tilde H &=& g^2 R^2 + k - \fft{\alpha^2 q^3 \tanh(\ft12\alpha^2 q\,u)
(3-\tanh^2(\ft12\alpha^2 q\,u))}{12R} + {\cal O}(R^{-3})\,.
\eea
Hence the spacetime is asymptotically AdS, flat or dS at large $R$ for $g^2>0$, $g=0$ and $g^2<0$, respectively.
The effective time-dependent ``Vaidya mass'' measured at infinity is given by
\be \label{timemass}
M(u)= \ft{1}{24}\alpha^2 q^3 \tanh(\ft12\alpha^2 q\,u)\big
(3-\tanh^2(\ft12\alpha^2 q\,u)\big)\ge 0\,,
\ee
which is a monotonically increasing function for $u\ge 0$ and approaches the final mass $M_0$ as $u\rightarrow \infty$. For general $u$, the metric has a power-law curvature singularity at $R=0$ which should be dressed by an apparent horizon at $R=R_0(u)>0$ defined by
\be
0=g^{\mu\nu} \partial_\mu R\, \partial_\nu R= \fft{\tilde H}{h^2}\,.
\ee
To identify the root structure of $\tilde H$ and related parameter ranges, one can follow a similar argument as in the static case. For instance, with $g^2>0$, an apparent horizon is guaranteed if $\tilde H(R=0)=k - \ft12\alpha^2q^2<0$. (For $g^2<0$, there can be a cosmological horizon in addition to a black-hole apparent horizon, of which the latter, for $q$ less than a critical value, is situated inside the former.) As $u$ goes to infinity, the apparent horizon will approach the event horizon.

The global structure at $u=0$ is more subtle. For small $u$, the scalar field and metric functions behave as
\be
\phi=\fft{\alpha^2q^2}{2R}u + {\cal O}(u^3)\,,\quad\tilde H=g^2 R^2 + k - \fft{\alpha^4 q^4}{8R}u + {\cal O}(u^3)\,,\quad
h=1 -\fft{\alpha^4 q^4 }{32 R^2}u^2 + {\cal O}(u^4)\,.
\ee
The polynomial curvature invariants are
\bea
R^\mu{}_\mu &=&-12g^2 -\fft{\alpha^4 q^4}{4R^3}u + {\cal O}(u^2)\,,\qquad R_{\mu\nu}R^{\mu\nu}=36g^4 + \fft{3g^2 \alpha^4 q^4}{2R^3} u + {\cal O}(u^2)\,,\cr
R_{\kappa\lambda\mu\nu}R^{\kappa\lambda\mu\nu} &=& 24 g^4 + \fft{g^2\alpha^4 q^4}{R^3} u + {\cal O}(u^2)\,.
\eea
At $u=0$ the spacetime becomes AdS, flat or dS except for a singular point at $R=0$ where the scalar polynomials are not well defined (their values depending on the path to $(0,0)$ in the $(u,R)$ plane). Nonetheless we can still connect the spacetime to an AdS, flat or dS vacuum for $u<0$ with $M(u)=0$. Up to the linear order in $u$, the energy-momentum tensor $T_{\mu\nu}\equiv R_{\mu\nu} -\ft12 (R +6g^2)g_{\mu\nu}$ (vacuum background subtracted) is given by
\be
T_{uu}=\fft{\alpha^4 q^4}{8R^3} - \fft{\alpha^4 q^4(g^2R^2+k)}{8R^3} u + {\cal O}(u^2)\,,\qquad
T_{ij} = \tilde g_{ij}\,\fft{\alpha^4 q^4}{8R} u  + {\cal O}(u^2)\,,
\ee
where $\tilde g_{ij}$ denotes the metric for $d\Omega_{2,k}^2$. (Note that $T^\mu{}_\mu = \alpha^4 q^4\,u/(4R^3) + {\cal O}(u^2)$, $T_{\mu\nu}T^{\mu\nu} = \alpha^8q^8 u^2/(16R^3) + {\cal O}(u^3)$, both of which vanish for $u=0$ and $R\neq 0$.) Thus the birth of our evolving black hole can be understood as a vacuum being kick-started by a singular energy-momentum tensor at $u=0$.  In this process, the singularity at $u=0=R$ may become globally naked if the initial mass accumulation is not fast enough (measured by $\lim_{u\to 0} M(u)/u$) \cite{Kuroda84}. We will discuss this further at the end of the paper.

We also point out that although it takes infinite $u$-time to reach the static limit, the function $\tanh(x)$ approaches 1 exponentially fast, i.e., $1-\tanh(x)\sim e^{-2x}$ as $x \rightarrow \infty$.  Thus we can define a characteristic relaxation time $u_0=1/(\alpha^2 q)\sim 1/(\alpha^4 M_0)^{1/3}$ such that the black hole reaches equilibrium at the rate of $e^{-u/u_0}$.  In addition, the mass (\ref{timemass}) behaves as $M(u)\sim  (1 -6 e^{-2u/u_0})M_0$.  It is clear that the bigger the mass $M_0$, the shorter the relaxation time $u_0$ becomes. This is a feature generally not seen in the Vaidya spacetime.

Furthermore, we present the solution in another two coordinate systems that may be instructive for future study. The first one is a diagonalized form. We introduce a new coordinate time $t$ and assume that the advanced time $u(t,R)$ satisfies $\partial u/\partial R = h/\tilde H$. This allows us to rewrite the metric as
\be
ds^2 = -(\partial_t u)^2 \tilde H dt^2 + \fft{h^2\,dR^2}{\tilde H} + R^2 d\Omega_{2,k}^2\,.
\ee
For $u\rightarrow 0$, we can make the following coordinate transformation
\bea
u &=& \tilde u + \fft{\alpha^4 q^4}{16R(g^2 R^2 + k)} \tilde u^2  + {\cal O}(\tilde u^3)\,,\qquad
\tilde u \equiv t + \fft{\arctan(gR/\sqrt{k})}{g\sqrt{k}}\longrightarrow 0\,,
\eea
so that the metric takes a Schwarzschild-like form
\bea
ds^2 &=& - \tilde f dt^2 + \fft{dr^2}{f} + R^2 d\Omega_{2,k}^2\,,\cr
\tilde f &=& g^2 R^2 + k + \fft{\alpha^4q^4}{8R} \tilde u  + {\cal O}(\tilde u^2)\,,\qquad
f = g^2 R^2 + k - \fft{\alpha^4q^4}{8R} \tilde u + {\cal O} (\tilde u^2)\,.
\eea
Thus we see that the evolution at early $u$ already indicates that the spacetime evolves to the scalar-hairy black hole rather than the Schwarzschild one.

Another coordinate system worth mentioning is when we use the dilaton $\phi$ itself as a coordinate to replace $r$.  The solution can be expressed as
\bea
ds^2 &=& \ft14{\rm csch}^2(\ft12 \phi)\Big[ 2\tanh(\ft12 \alpha^2 q\,u) d\phi\, du
- \hat H du^2 + q^2\tanh^2(\ft12\alpha^2 q\, u)\,d\Omega_{2,k}^2\Big]\,,\cr
\hat H &=& -4k\sinh^2(\ft12\phi) -
g^2q^2\tanh(\ft12\alpha q\,u)+\alpha^2 q^2\big(\sinh\phi - \phi\,\tanh^2(\ft12\alpha^2 q u)\big)\,.
\eea
By this means, the infinity is now located at $\phi=0$, and the dilaton does not appear to evolve with the ``time'' $u$.

{\bf Generalized Roberts solution:} We now consider the special case of gauged supergravity with $\alpha=0$.  To extract a non-trivial solution in this limit, we let $q\rightarrow \sqrt{2q}/\alpha$ and send $\alpha\rightarrow 0$; then we find
\be
ds^2 = 2du dr - \big(k-q + g^2r(r+ qu)\big) du^2 + r(r + q u) d\Omega_{2,k}^2\,,\qquad e^\phi = 1 + \fft{q u}{r}\,.\label{al0}
\ee
The solution has two singularities, one located at $r=0$ and the other at $u=\infty$.
With the luminosity coordinate $R$, the metric is
\be
ds^2 = \fft{4R\, du dR}{\sqrt{4R^2 + q^2 u^2}} - \Big(g^2R^2 + k - \fft{q^2 u}{\sqrt{4R^2 + q^2 u^2}}\Big) du^2 + R^2 d\Omega_{2,k}\,.
\ee
If we further set $g^2=0$, the theory (\ref{lag1}) simply reduces to Einstein gravity coupled to a free scalar field, and for (\ref{al0}) the two-dimensional metric on the $(u,r)$ plane becomes flat.  This special case was first obtained and analysed by Roberts \cite{Roberts89}.  Our version therefore provides a supergravity or (A)dS generalization and may be further examined in the context of critical phenomena \cite{Gundlach07}, cosmic censorship and cosmology .

{\bf Concluding remarks}:  In this paper, we considered Einstein gravity coupled to a dilatonic scalar with the scalar potential (\ref{lag1}).  The potential contains two non-trivial parameters, the asymptotic cosmological constant $\Lambda=-3g^2$ and $\alpha^2$ that is responsible for the existence of a static scalar-hairy black hole instead of a Schwarzschild one.  The $g^2$-term arises in $D=4$, ${\cal N}=4$ gauged supergravity.  We obtain an exact solution that describes the formation of the scalar hairy black hole in an (A)dS or flat background.  At the initial time, the solution can be smoothly connected to a suitable vacuum except one point, which is singular owing to a singular energy-momentum tensor that kick-starts the time evolution.  A naked singularity may occur as the solution approaches the static limit, in the exponential fashion $e^{-u/u_0}$ with $u_0=1/(\alpha^4 M_0)^{1/3}$.  Hence a larger initial source will result in a shorter $u_0$ and a bigger black hole mass $M_0$.

To be more specific, let us consider the case $g^2\sim 0$ and $k=1$. The existence of an event horizon requires that $q>\sqrt{2}/\alpha$, and all black holes must have $M_0\gtrsim 1/\alpha$. Meanwhile, the event horizon of the static black hole is located at $r_+\sim M_0$; for the black hole formation, we therefore have the relaxation time $u_0\lesssim r_+$ and the information of singularity will be trapped. For $M_0\lesssim 1/\alpha$, a black hole cannot be formed and the naked singularity persists.  However, if $\alpha\sim 1/\ell_p$, where $\ell_p$ is the Planck length, then the corresponding Compton wavelength $\ell_{\rm comp}=\ell_p^2/M_0$ will be larger than $r_+$ and hence the singularity will be smeared out by quantum effect.

To summarize, we provide a first example of exact black hole formation from a Lagrangian of fundamental fields. The asymptotic spacetime is either (A)dS or flat. This flexibility can facilitate various future applications, such as modelling real life black hole and investigating thermalization of certain strongly coupled field theory.

\section*{Acknowledgement}

We are grateful to Michal Artymowski, Daniel Finley, Sijie Gao, Hai-shan Liu, Chris Pope, Zhao-Long Wang for useful discussions. The research is supported in part by the NSFC grants 11175269 and 11235003.

\end{document}